\documentclass[a4paper,fleqn,usenatbib]{mnras}

\usepackage{microtype}

\usepackage[T1]{fontenc}
\usepackage{ae,aecompl}

\usepackage{graphicx}	
\usepackage{amsmath}	
\usepackage{amssymb}	

\title[Binary origin of a CCO]{Evidence for a binary origin of a central compact object.}

\author[V. Doroshenko et al.]{
Victor Doroshenko$^{1\ast}$, Gerd P{\"u}hlhofer$^{1}$, Patrick Kavanagh$^{1,2}$, Andrea Santangelo$^{1}$, 
\newauthor
Valery Suleimanov$^{1,3}$, Dmitry Klochkov$^{1}$\\
\\
\normalsize{$^{1}$IAAT, University of Tuebingen, Sand 1, Tuebingen, 72076, Germany}\\
\normalsize{$^{2}$School of Cosmic Physics, Dublin Institute for Advanced Studies, 31 Fitzwillam Place, Dublin 2, Ireland}\\
\normalsize{$^{3}$Kazan (Volga region) Federal University, Kremlevskaja str., 18, Kazan 420008, Russia}\\
\normalsize{$^\ast$E-mail:  doroshv@astro.uni-tuebingen.de.}
}

\date{Accepted XXX. Received YYY; in original form ZZZ}

\pubyear{2015}

\begin{document}
\label{firstpage}
\pagerange{\pageref{firstpage}--\pageref{lastpage}}
\maketitle


\begin{abstract} Central compact objects (CCOs) are thought to be young
thermally emitting isolated neutron stars that were born during the preceding
core-collapse supernova explosion. Here we present evidence that at
least in one case the CCO could have been formed within a binary system. We
show that the highly reddened optical source IRAS~17287$-$3443, located
$25^{\prime \prime}$ away from the CCO candidate XMMUJ173203.3$-$344518 and
classified previously as a post asymptotic giant branch star, is indeed
surrounded by a dust shell. This shell is heated by the central star to temperatures
of $\sim90$\,K and observed as extended infrared emission in 8-160\,$\mu$m
band. The dust temperature also increases in the vicinity of the CCO which
implies that it likely resides within the shell. We estimate the total dust mass
to be $\sim0.4-1.5\,M_\odot$ which significantly exceeds expected dust yields
by normal stars and thus likely condensed from supernova ejecta. Taking into
account that both the age of the supernova remnant and the duration of active
mass loss phase by the optical star are much shorter than the total lifetime of
either object, the supernova and the onset of the active mass loss phase of the
companion have likely occurred approximately simultaneously. This is most easily
explained if the evolution of both objects is interconnected. We conclude,
therefore, that both stars were likely members of the same binary system
disrupted by a supernova. \end{abstract}

\begin{keywords}
stars: AGB and post-AGB -- stars: neutron -- binaries: general -- stars: formation -- ISM: supernova remnants
\end{keywords}



\section{Introduction}

Central compact objects (CCOs) are X-ray point sources without optical, radio,
or pulsar wind nebula counterparts, found close to the centres of several young
supernova remnants (SNRs). Their X-ray fluxes are
constant within observational constraints and potential counterparts that would
hint at accretion-dominated emission scenarios are not known. X-ray pulsations
have been detected from three out of the eight CCOs known so far. All this
suggests that CCOs are likely young $\le10^4$\,yr isolated neutron
stars (NSs), still cooling after their birth in a core-collapse supernova explosion
\citep{Pavlov04, deLuca08}. Measurements of the spin-down rates of pulsating CCOs imply
very low dipole magnetic fields of the order of $10^{10}-10^{11}$\,G, compared
to the regularly measured neutron star magnetic fields of $\sim
10^{12}\,\mathrm{G}$ or to the even higher magnetic fields derived for
magnetars. For this reason CCOs are sometimes dubbed ``anti-magnetars''
\citep{Halpern10,Gotthelf13}. It is unknown, however, whether all of the CCOs
are weakly magnetised, and whether they were born with these low magnetic
fields.

In fact, detailed analysis of individual sources reveals CCOs as a rather
diverse and intrinsically controversial sample. The high pulsed fractions
observed in pulsating CCOs imply strong magnetic fields are necessary to
explain non-uniform cooling of the NS surface \citep{Bogdanov14}. The long spin
period (6.7\,h) of the CCO candidate 1E\,161348$-$5055 \citep{Garmire00} and
superluminal echoes detected from Cas\,A \citep{Krause05} also favour
magnetar-like CCO fields at those objects.
Recently, \cite{Ho11} proposed that the dipole magnetic field of CCOs could have been ``buried'' during a hyper-accretion episode shortly after the
supernova explosion. In this case different observational appearances of
individual objects could be attributed to different properties of their birth
environment.

Here we report the discovery of extended infrared emission towards the centre
of the non-thermal shell-type SNR~G353.6$-$0.7 which hosts the CCO candidate
XMMUJ173203.3$-$344518. We show that the observed emission is associated with
dust heated by the optical source IRAS~17287$-$3443, previously classified by
\cite{Suarez06} as a post asymptotic giant branch (post-AGB) star. 
The large mass of dust in the shell suggests that the post-AGB wind is carving the
proto-planetary nebula from material produced in the SN explosion and further
supports the association with the SNR.

Both the SNR age and the duration of the active mass loss phase of
the central star are much shorter than the lifetime of both objects, suggesting
that the supernova explosion and the mass loss by the central star occurred
approximately simultaneously. Such coincidence suggests that the evolution of
two objects is likely interconnected which can only happen if both objects were members of the same
binary system disrupted by the supernova explosion. We note that in this case
the ``buried'' magnetic field scenario at the CCO is more likely to be realised
as otherwise accreted material could only be supplied from the supernova ejecta.

\section{Extended infrared emission}
In this study, we used data products provided as part of the GLIMPSE
\citep{Glimpse} and MIPSGAL \citep{Mipsgal} surveys performed with
the Infrared Array Camera (IRAC, 3.6-8\,$\mu$m) and the Multi-band Imaging
Photometer (MIPS, 24-160\,$\mu$m) onboard the Spitzer space telescope. We also
used public data (observations 1342214713, 1342214714) from the Photodetector
Array Camera \& Spectrometer (PACS, 70-160\,$\mu$m) onboard the HERSCHEL space
telescope, and XMM-Newton observations (0405680201, 0673930101, 0694030101,
0722090101, 0722190201) to image the SNR X-ray shell in the 0.2-10\,keV energy
range. Finally, we used Chandra ACIS-I data (observation 9139) for point source
localisations. In all cases, we followed standard reduction procedures
described in the respective instrument's documentation.

\begin{figure}
	\centering
	\includegraphics[width=0.49\textwidth]{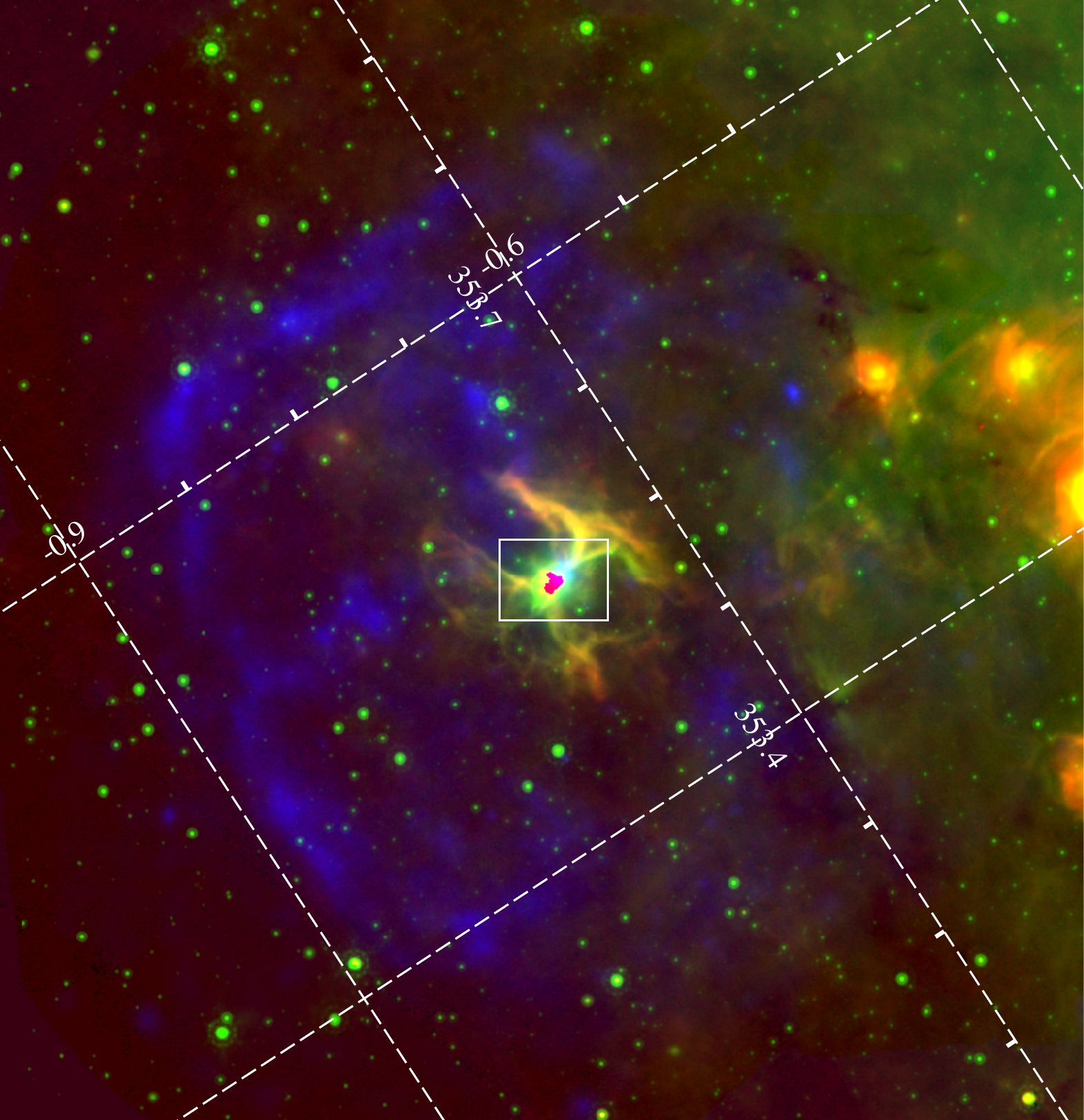}
	\caption{False-Colour X-ray and infrared emission image in the region around
	the SNR G353.6$-$0.7. The RGB colours correspond to HPACS 70\,$\mu$m (red), MIPS
	24\,$\mu$m (green), and XMM-Newton 0.2-10 keV (blue) data, respectively.
	The intensity scale is logarithmic for all channels. The SNR shell in X-rays and the infrared dust shell are visible in the centre. The small box indicates the central part of the structure presented in Fig~\ref{fig:closeup}
	in more detail. Galactic coordinates are also shown for reference.}
	\label{fig:overview}
\end{figure}

As part of an ongoing multi-wavelength analysis of the SNR emission, we
inspected MIPSGAL maps covering the source. Immediately, the very peculiar
extended structure coincident with the geometric centre of the X-ray SNR shell
attracted our attention (see Fig.~\ref{fig:overview}). Extended emission is
detected across the 8\,$\mu$m to 160\,$\mu$m wavelength range, also in WISE and
AKARI all-sky surveys. The highest luminosity is seen in the MIPS 24$\mu$m
band. The extended emission region has a radius of $\sim0\farcm5-2\farcm5$, and
its surface brightness strongly peaks towards the bright inner $0\farcm5$ core,
particularly at shorter wavelengths. In the IRAC 8\,$\mu$m band, only the core
is detected, and no extended emission is detected at shorter wavelengths. This
strongly suggests that the warm dust with a temperature of $\sim90$\,K heated
by a central source is responsible for the observed emission. We note that
while collisional heating by hot shocked ejecta is normally expected to be an
important ingredient in the thermal balance of dust residing within SNRs
\citep{Dwek10}, it is apparently not the case for G353.6-0.7.

Indeed, the temperature of the collisionally heated dust is expected to be
defined by the temperature and density of hot ejecta and be fairly uniform
throughout the SNR. However, as illustrated in Fig.~\ref{fig:sed}, the observed
dust temperature radial profile of the bright central shell is fully consistent
with heating by the central source. Moreover, the observed X-ray emission from
the remnant is known to be purely non-thermal and thus provides no
observational evidence for presence of hot ejecta in this source
\citep{Acero09,Bamba12}. We conclude, therefore, that heating by the central
source dominates the dust thermal balance in the core of the shell. We note,
however, that the heating efficiency of the central star decreases rapidly with
distance, so collisional heating might still play some role in the outer parts of
the remnant.

\begin{figure}
	\centering
	\includegraphics[width=0.49\textwidth]{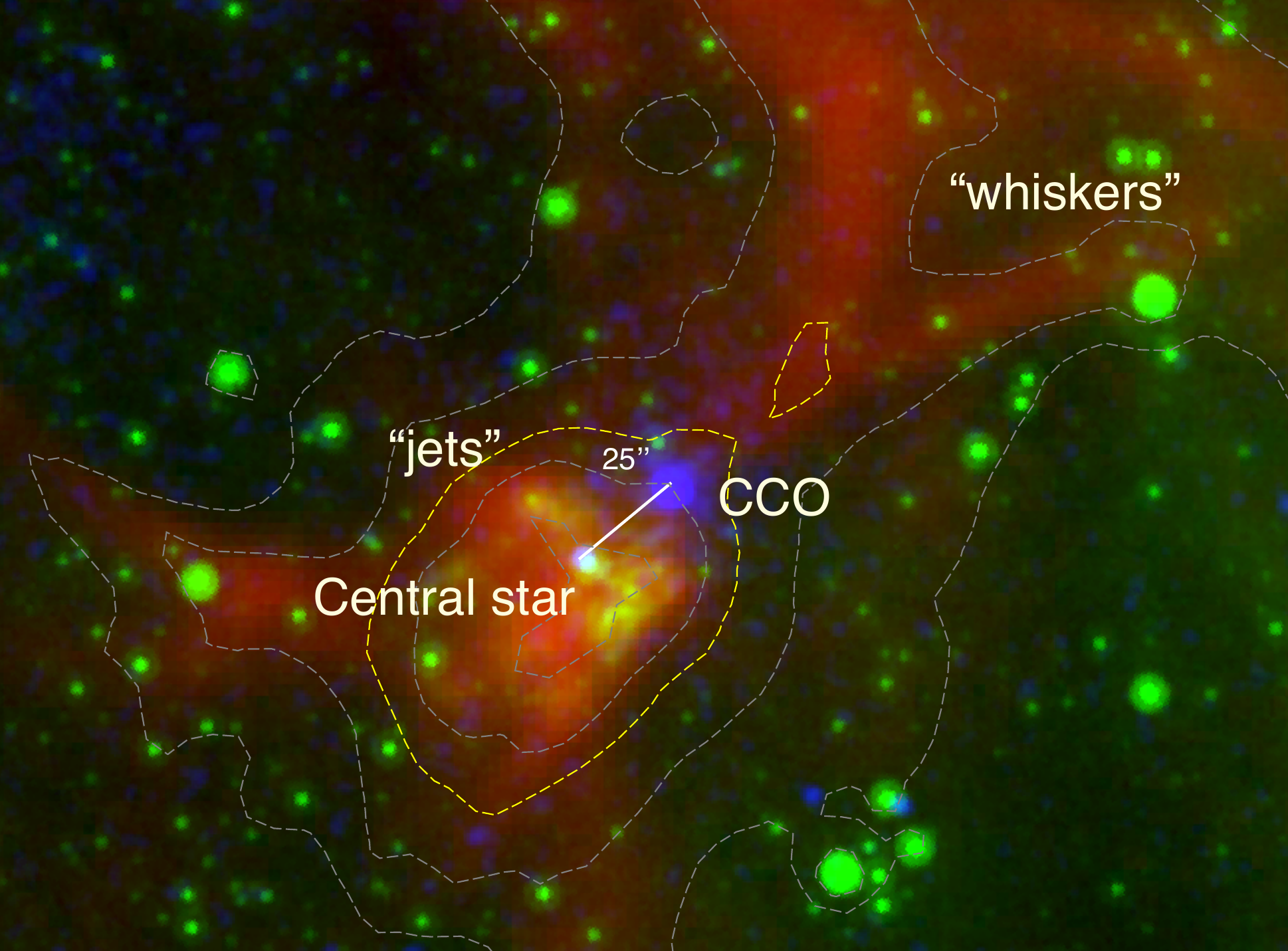}
	\caption{False-Colour X-ray and infrared emission image from the core of
the infrared shell. The RGB colours correspond to Chandra X-ray 0.2-10 keV (blue),
IRAC infrared 8\,$\mu$m (green), and HPACS 70\,$\mu$m (red) data. The intensity scale is logarithmic for all channels. Overlaid
are equal brightness levels from the MIPS 24\,$\mu$m band. Note that around the CCO the infrared emission is suppressed in the 70\,$\mu$m band and enhanced in the 24\,$\mu$m band suggesting higher dust temperature.}
	\label{fig:closeup}
\end{figure}

A bright point source only slightly offset
($\sim10^{\prime\prime}$) from the apparent geometrical centre of the extended
structure is  detected in the near-infrared, optical, ultraviolet (UV),
and X-ray bands. We label it the ``central star'' (the object should not be
confused with the CCO). It is the only bright source close to the centre of the
extended emission and thus is likely the heating source for the dust. Analysis
of the combined spectral energy distribution (SED) of the central star and the
dust shell (see section~\ref{sec_sed}) and of the observed dust's radial
temperature profile (see Fig.\,\ref{fig:sed} and appendix~\ref{dustmass})
strongly support this conclusion.

Two jet-like structures are clearly visible at 8\,$\mu$m, extending from the
central star in the direction perpendicular to the line connecting the star
with the CCO, as illustrated in Fig.~\ref{fig:closeup}. These ``jets'' can also
be traced at longer wavelengths. A second axis of symmetry, bisecting the outer
``whiskers'' visible in the 24 and 70\,$\mu$m bands in the southeast-northwest
direction, also crosses the central star, as seen in Fig.\,\ref{fig:closeup}.
We conclude, therefore, that the morphology of the extended emission also
supports an association with the central star.

Indeed, the CCO and the dust shell appear to be interacting. In particular, the
infrared emission around the CCO is suppressed in the 70\,$\mu$m band and
enhanced in the 24\,$\mu$m band as illustrated in Fig.~\ref{fig:closeup}. This
strongly suggests additional heating of the dust by X-ray emission from the
neutron star which can only be effective if the neutron star is submerged
in the dust shell. The faint X-ray glow around the CCO, as visible in
Fig.~\ref{fig:closeup}, has been shown to be inconsistent with pure
point-source emission and indeed was interpreted by \citep{Halpern10a} as a
dust-scattering halo around the CCO. We conclude, therefore, that the dust
shell responsible for the observed infrared emission encloses both the CCO and
the central star which heats the dust to the observed temperatures, and thus all
three objects reside within the SNR shell. In the remaining sections we will
demonstrate that all observed properties of the dust and the central source are
consistent with this conclusion.

\subsection{SED analysis}\label{sec_sed} 
\begin{figure*}
	\centering
		\includegraphics[width=0.49\textwidth]{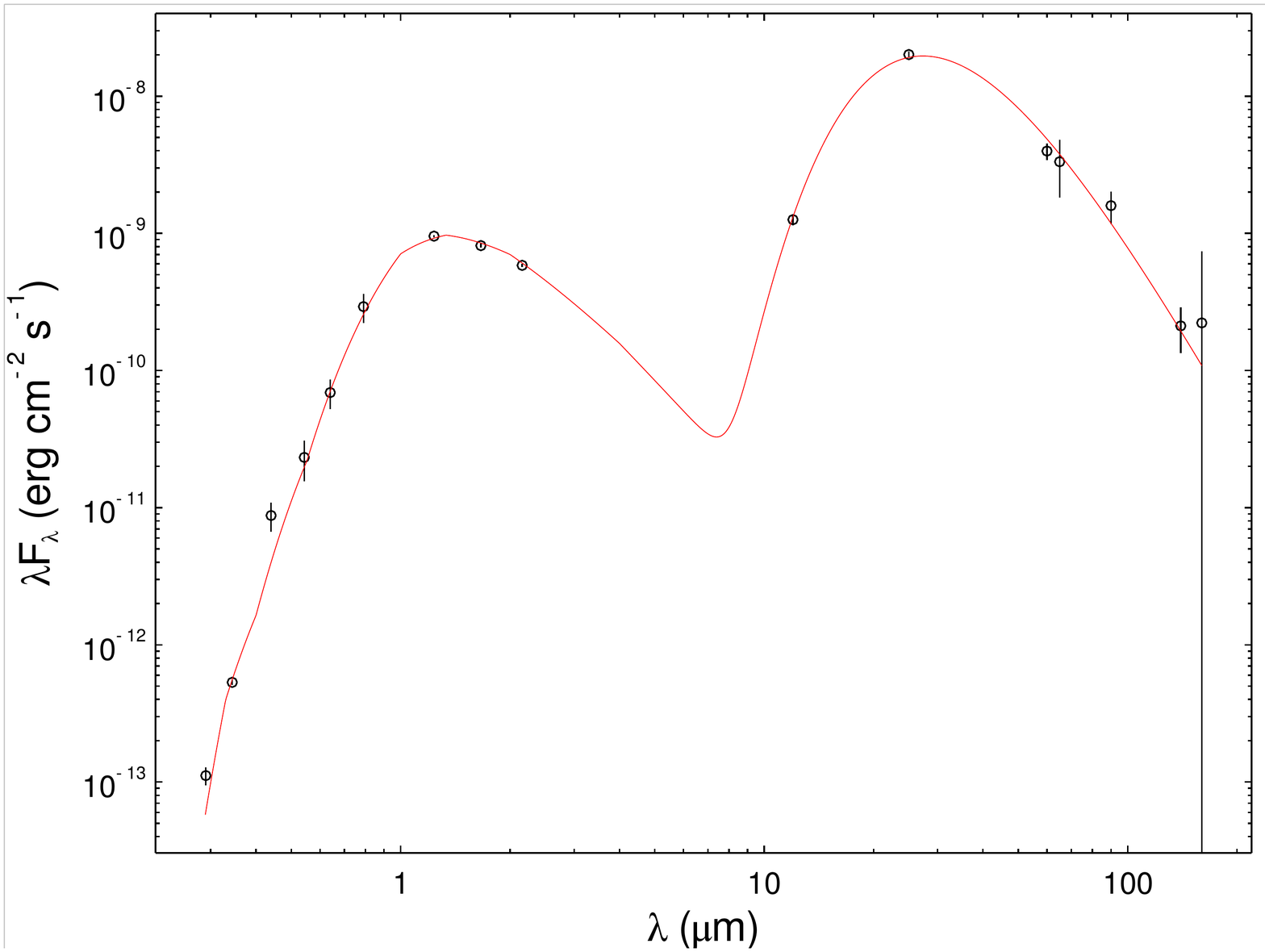}
		\includegraphics[width=0.49\textwidth]{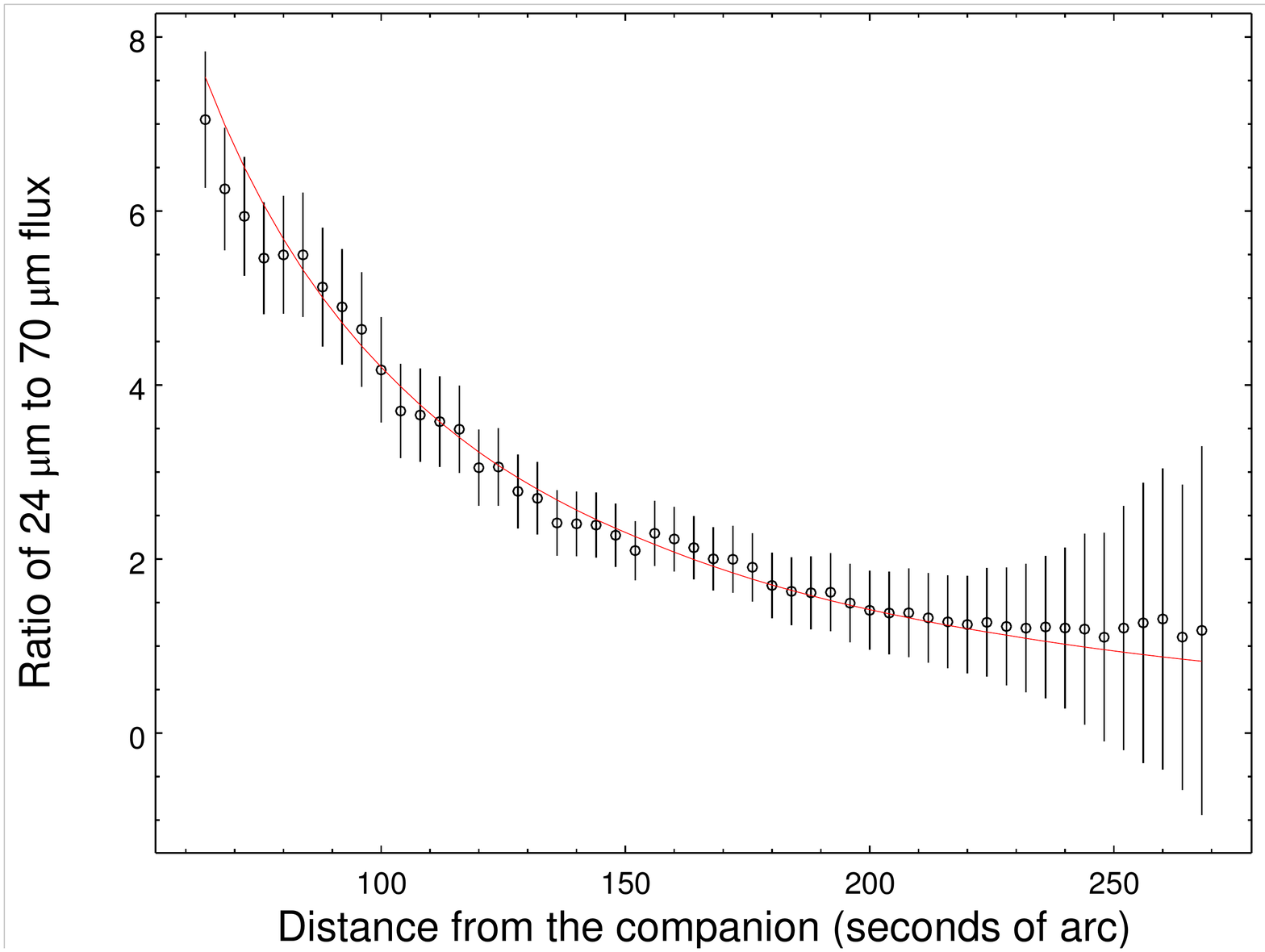}
		\includegraphics[width=0.49\textwidth]{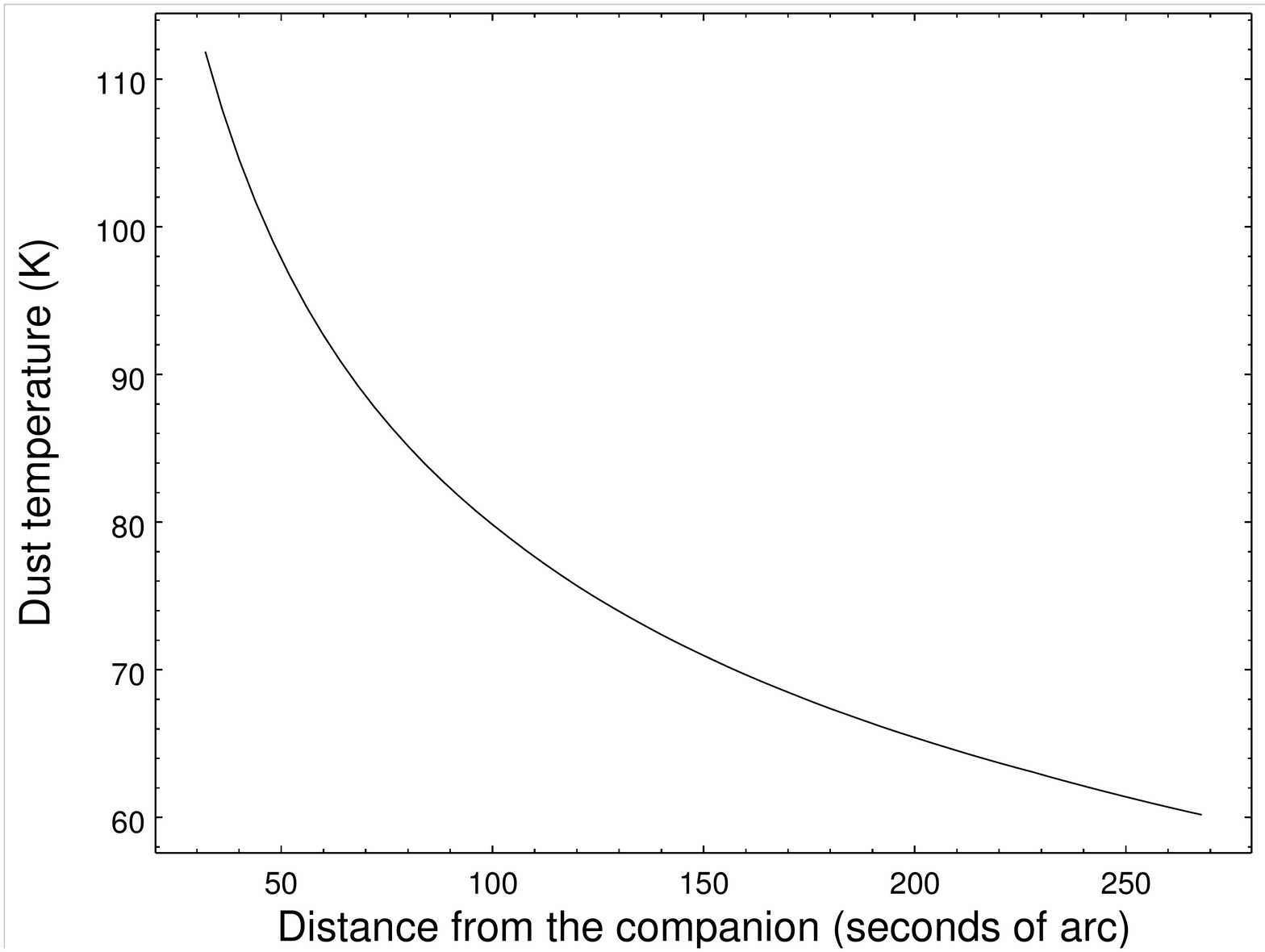}
		\includegraphics[width=0.49\textwidth]{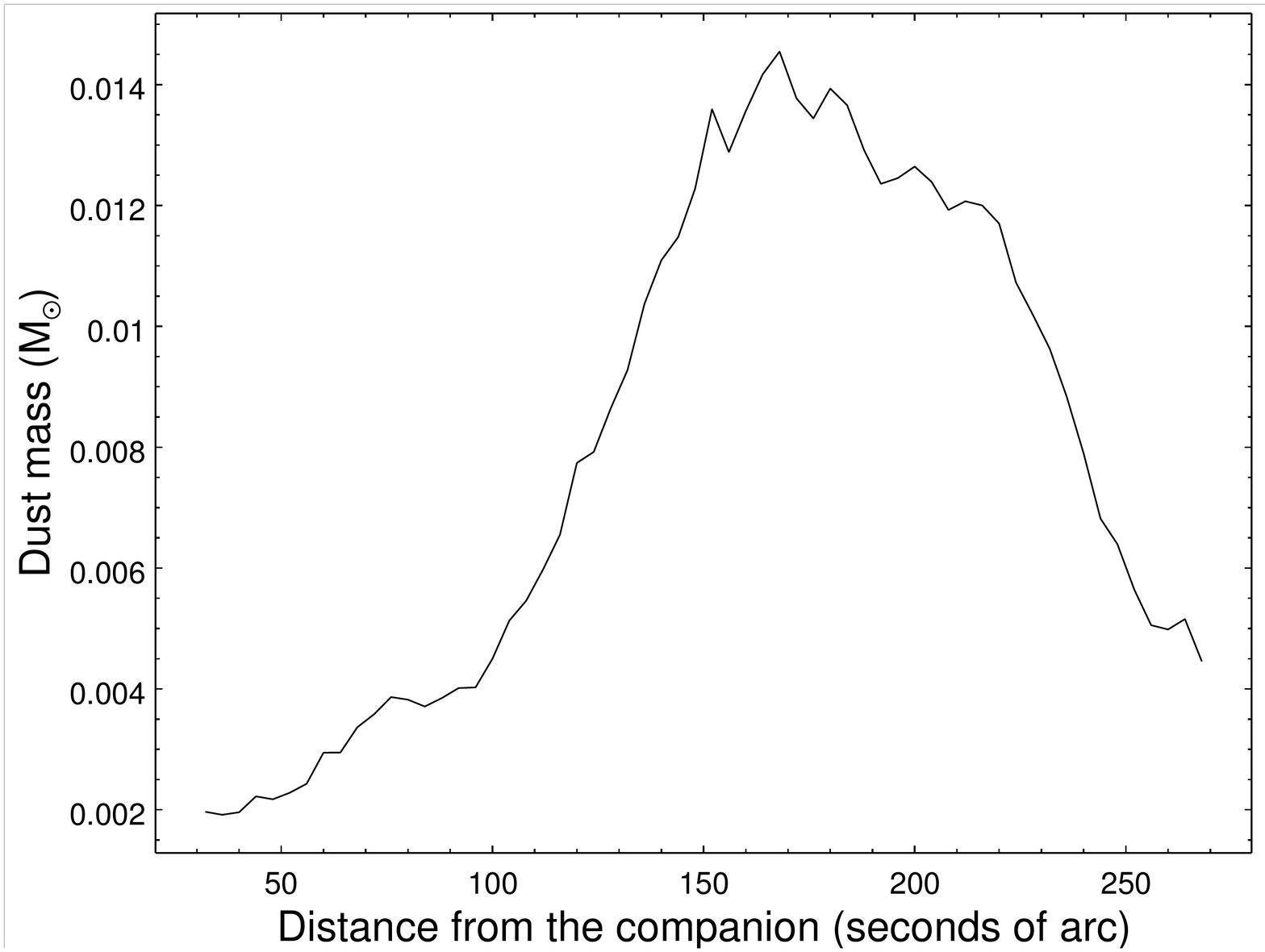}
		\caption{Top left: Combined broadband-SED of the central post-AGB star and its dust shell.
		Top right: Observed ratio of 24 to 70\,$\mu$m fluxes as a function of the distance 
		to the central star. The resulting radial
		distribution of the temperature (bottom left) and projected dust mass (bottom right) in the
		shell, estimated from the observed 24 and 70\,$\mu$m fluxes, are shown for an assumed
		grain size of $a = 0.1$\,$\mu$m.}
	\label{fig:sed}
\end{figure*}

To obtain the SED of the central star, we used optical and near-infrared fluxes
reported in the USNO-B \citep{usno_b} and 2MASS \citep{twomass} catalogues. In
addition, data from the optical monitor onboard the XMM-Newton observatory were
used to estimate fluxes in the ultraviolet band. The dust emission is dominated
by the compact core and appears as a point source for most instruments, so as
as a first approximation for the SED analysis of the extended emission, we used
calibrated fluxes reported for IRAS~17287-3443 in the AKARI-FIS
\citep{akari_fis}, the IRAS-PSC \citep{iras_psc}, the MSX \citep{msx_psc}, and
the WISE-PSC \citep{Wright10} catalogues. The combined SED for both the central
star and the dust shell is presented in Fig.~\ref{fig:sed}.

To describe the observed SED, we assumed blackbody-type emission subject to
interstellar extinction \citep{Fitzpatrick99} for the central star, whereas for
the dust emission we used a modified blackbody model \citep{Draine84} of the
form $F_\lambda\sim\lambda^{-\beta}B_\lambda(T_{\mathrm{dust}})$. Here $\beta$ is the
dust emissivity index and $B_\lambda(T_{\mathrm{dust}})$ is the blackbody photon
distribution (see also appendix~\ref{dustmass} for details). The two components
can be fit independently with best-fit temperatures of $T_*\ge 9000$\,K for the
optical star and $T_{\mathrm{dust}}\simeq90$\,K for the dust with an assumed typical
grain size of 0.1$\mu$m and $\beta = 2$ \citep{dust_beta}. 

Strong intrinsic correlation between the temperature and the extinction
coefficient together with high foreground absorption towards the source make
the temperature (and thus luminosity) of the central star unconstrained from
the fit even though the SED contains UV fluxes. This degeneracy can be resolved
with additional assumptions on the extinction in the direction of the source. For
instance, the extinction coefficient $A_V\sim9$ can be estimated based on the
observed X-ray absorption in the direction of the CCO \citep{Acero09, Tian10,
Halpern10a, Klochkov13}, as detailed in appendix~\ref{xraynh}. This implies
the luminosity of $\sim6\times10^3\,L_\odot$ which, as discussed below, is
sufficient to heat the dust to the observed temperature.
 
Comparison of the 8, 24, and 70\,$\mu$m images implies that
the dust temperature increases towards the centre of the shell, thus strongly
suggesting heating by the central source. Modelling of the dust temperature or,
equivalently, the observed flux ratio in the 24 and 70\,$\mu$m as function of
distance from the central source provides, therefore, another way to constrain
its luminosity as described in appendix~\ref{dustmass}. In fact, simultaneous
modelling the SEDs of the optical star and the dust together with the observed
radial temperature profile (via the flux ratio in 24 and 70\,$\mu$m bands)
allows us to unambiguously relate the temperature of the central source and the
dust grain size which remains the only unknown parameter. Note that in this
case no assumptions on the extinction coefficient are required. However, we note
that its best-fit value remains consistent with the estimate based on X-ray
absorption for an assumed average grain size $a\sim0.1\mu$m, typical for cosmic
dust. The results for this case are presented in Fig.~\ref{fig:sed} and
Table~\ref{tab:sed}.

Note that, as we show in section \ref{dustmass1}, the temperature and the
luminosity of the central star are consistent with both the
observed SED (under the assumption that the source is at the same distance as the
SNR) and with the previous classification of the central source as a post-AGB core
by \cite{Suarez06} although detailed optical spectroscopy is required to
unambiguously establish its spectral class. Therefore, we conclude that the SEDs of
the dust and the central source are consistent with the hypothesis that
the observed infrared emission comes from the dust shell heated by the central star
located within the SNR.

\begin{table}
	\begin{center}
	\begin{tabular}{cccccc}
		$A_V$& $T_*/1000$\,K & $L_*/L_\odot$  &  $T_{\mathrm{dust}}$& $M_{\mathrm{dust}}/M_\odot$& $\beta$\\
		\hline
		8.67(3) &91(5) & 5788(25) & 89(2) & 0.4(1) & 2.0(1)\\
	\end{tabular}
	\end{center}
	\caption{Best-fit parameters for the joint fit of the SED and radial temperature
	profile of the shell for an assumed grain size $a = 0.1\,\mu$m and ignoring projection effects. Uncertainties to last significant digit at $1\sigma$ confidence level are given in the parenthesis.}
	\label{tab:sed}
\end{table}

\section{Discussion} 
\subsection{Total dust mass}\label{dustmass1} To understand the origin of the
observed dust it is important to estimate its mass. In principle, for
single-temperature dust its mass can be estimated directly from the normalisation
of the SED. However, given the observed temperature gradient away from the
central star, a single temperature model is not justified. Indeed, the strong
dependency of dust emissivity on temperature $L\sim <Q_{IR}>T^4$ implies that
the colder outer regions do not contribute much to the observed flux, but might
contain a large fraction of the total dust mass (here $<Q_{IR}>$ is
the Planck-averaged dust emissivity for a given temperature). Therefore, we
estimated the mass of the shell using the radial mass profile derived using
observed radial flux profiles in the 24\,$\mu$m and 70\,$\mu$m bands under the
assumption that the dust is in thermal equilibrium with the irradiating flux at
each distance from the central star. The resulting projected dust mass profile
for $a = 0.1\,\mu$m grains is presented in Fig.~\ref{fig:sed}. Note that due to
projection effects this figure may underestimate the total mass by up to an
order of magnitude (see appendix~\ref{dustmass}).

The assumed grain size also strongly affects the estimated mass of the dust. In
particular, grain sizes from $a = 1.5\,\mu$m to $a = 0.001\,\mu$m correspond to
total dust masses between 0.03 and 50~$M_\odot$. Such a large uncertainty
appears mainly because the SED of the central source does not allow us to
constrain its reddening, effective temperature, and luminosity simultaneously.
This degeneracy can be resolved through constraints on the central star's
luminosity, either observational or theoretical. For instance, as discussed
above, optical extinction can be estimated based on the observed X-ray
absorption which would result in shell parameters close to those reported in
Table~\ref{tab:sed}.

\begin{figure}
	\centering
		\includegraphics[width=0.49\textwidth]{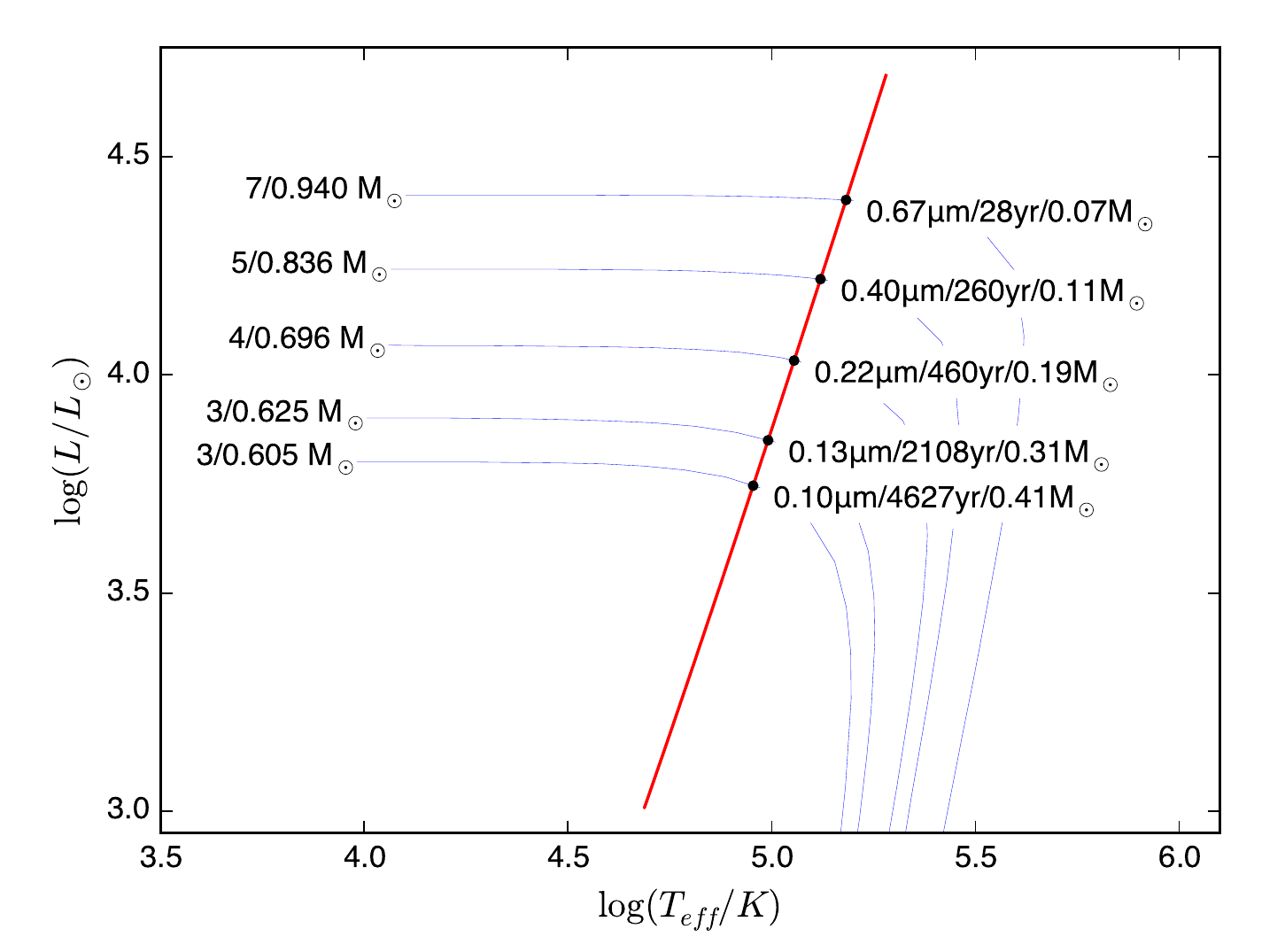}
	\caption{The temperature of the central source as derived from a combined SED and
temperature profile analysis (red line), and evolutionary tracks of post-AGB
stars with different masses. The dust grain size, the age of the star, and the total mass of
the dust shell (for an assumed distance of 3.2\,kpc) corresponding to each track are also shown.}
	\label{fig:lumage}
\end{figure}

Furthermore, temperatures and luminosities of real stars are not arbitrary and
can be estimated based on stellar evolution considerations. For instance, if we
assume that the central star is indeed going through a post-AGB phase, the
luminosity can be constrained using the theoretical evolutionary tracks of
post-AGB stars as presented in Fig.~\ref{fig:lumage} for several
representative initial masses \citep{Bloecker95}. In the same figure, the
temperature and luminosity of the central source required to explain the
observed dust temperature profile is shown (based on the joint fitting of the
SED and dust temperature profile as function of grain size acting as a free
parameter). 

Stars evolve along the tracks toward higher temperatures so the
intersection of this line with evolutionary tracks not only implies a certain
stellar luminosity, but also the time since the beginning of the AGB phase. 
Note that stars with large initial masses evolve too fast to survive long
enough to explain both the observed infrared shell temperature and extent, so
low luminosity tracks are preferred. This conclusion is also consistent with
the relatively low observed X-ray luminosity of the central star (assuming the same
X-ray to optical flux ratio as in other post-AGB stars
\citep{Ramstedt12_xray_to_optical} of $10^{-5}$). We would like to emphasise
that evolutionary tracks with lower luminosities in this case also correspond
to dust with a canonical grain size of 0.1\,$\mu$m. 

Note that different assumptions on distance to the source would not move the
red line in Fig.~\ref{fig:lumage} since these do not affect the dust
temperature radial profile (see appendix~\ref{dustmass1}). On the other hand,
the derived dust mass and grain size do change, increasing and decreasing with the
assumed distance for given luminosity, respectively. We conclude, therefore, that while the
uncertainty in distance remains, both the SED of the central star and the dust
temperature radial profile can be consistently modelled if dust consisting of
$\sim 0.1\,\mu$m grains (which is typical for dust observed in the Galaxy) is
heated by a low mass post-AGB star with age comparable to the age of the SNR.

This implies, however, that the dust must consist of grains with the average size
of $0.1$\,$\mu$m, which corresponds to a total mass of dust in the shell of
$\sim0.4\,M_\odot$ ignoring projection effects (and up to 1.5-3\,$M_\odot$
if these are taken into account, see appendix~\ref{dustmass}). This is at least
two orders of magnitude more than dust yields expected from AGB stars which are
between $10^{-5}-10^{-2.5}\,M_\odot$
\citep{Meijerink03_agb_mass_loss,Ventura12}, or, in fact, from any other known
stars. There are several possibilities to explain this discrepancy.

First, the mass of the shell could be overestimated if the assumption that the dust
is only heated by the central source does not hold. Inspection of the
8-70\,$\mu$m images does indeed suggest that there is some emission in
the North-Eastern and South-Western parts of the remnant outside of the inner
dust shell, which seems to be spatially correlated with the diffuse X-ray
emission from the SNR. Therefore, it is unlikely to be background emission. It
also can not be dust heated by the central star due to large spatial distance
separating them. In fact, such emission might be what one normally would expect
to see within an SNR where the dust condensed from supernova ejecta is
collisionally heated by cooling ejecta gas. However, the extended structure
considered in our analysis is much brighter than anything else within the
remnant and has a temperature distribution fully consistent with heating by a
point source which thus dominates thermal balance of the dust.

On the other hand, past studies of dust formation in post-AGB atmospheres
assumed that the dust condenses from the wind of an isolated star, expanding
into the ISM. Apparently, this is not true for a star which was previously a
member of a binary system which now resides inside the boundaries of an SNR
containing metal-rich supernova ejecta. Indeed, conditions in winds of post-AGB
stars are much less severe than those in supernova ejecta and are more suitable
for dust condensation. However, atmospheres of AGB stars mostly consist of
hydrogen which leads to comparatively low dust yields. Therefore, mixing of
post-AGB wind with metal-rich material from supernova ejecta is likely to
enhance overall dust production.

Finally, the dust contained in the proto-planetary nebula may not at all
originate from the central star. Typically, the dust in proto-planetary nebulae
has been pre-supplied during the intense mass-loss phase of the AGB star
\citep{Kwok}. However, SNe themselves are also prodigious formers of dust with
masses of $>0.5$~M$_{\odot}$ possible in core-collapse ejecta, see e.g.
\cite{Bianchi}. Typically, the passage of the SNR reverse shock will destroy
most, if not all, of the ejecta dust, \citep{Noz2007}. However, in the case of
SNR G353.6$-$0.7, which is a purely non-thermal remnant in X-rays
\citep{Hess_discovery}, we would not expect this to happen since non-thermal
SNRs are thought to be evolving in very low-density bubbles blown by the
progenitor massive star \citep{Ber2010}, prohibitive to the formation of a
strong reverse shock. Therefore, the majority of the dust could have survived
to the current age. Assuming that wind from the central star has sufficiently
high velocity, it can catch and shape some of this dust. Taking into account
that the spatial size of the remnant exceeds that of the dust shell just by
factor of three, this does not seem improbable. This scenario would offer
further evidence for the association of the CCO and the central star. Note that
in this case the central star is, in principle, not required to be an effective
dust producer and must only have a sufficiently fast wind and high luminosity.
These requirements are satisfied, for instance, by most WR stars. The
classification of the central source as a post-AGB star by \cite{Suarez06}
seems, however, quite robust, and there is no apparent reason to question it
even if the bulk of the dust is not condensed directly from stellar material.

\subsection{Chance neighbours or relatives?} As discussed above, all
observational facts point to the conclusion that the dust shell encloses both
the CCO and the central star. Adopting the distance to the SNR of 3.2\,kpc
\citep{Hess_discovery, Tian10,Klochkov13,Klochkov15}, the observed angular distance of
$\sim25^{\prime\prime}$ between the CCO and the central star implies a
projected distance of just $\sim0.4$\,pc between the two objects\footnote{
Distances well beyond the Galactic centre are strongly disfavoured since they would imply an unrealistically large CCO radius 
\citep{Klochkov13,Klochkov15}, and also a TeV luminosity of the SNR by far exceeding that of other similar sources \citep{Acero15}. A distance at 5-6 kpc \citep{Fukuda14} would not change our conclusions.}.
The average density of AGB-stars in the Galaxy is $\sim 1\mathrm{\,kpc}^{-3}$ \citep{Jackson02}, so the probability that the two objects are 
chance neighbours is very low. 
Given the nature and the recent history of the two objects,
the hypothesis that they were members of the same binary system before the
supernova explosion of the CCO's progenitor star is immediately very suggestive.

This conclusion is indirectly supported by the observed morphology of the
infrared emission. Indeed, the jet-like structures seen at shorter wavelengths
form a clear axis of symmetry resembling the one seen in bipolar planetary nebulae
\citep{Dobrinic08}, which are usually associated with collimation of outflowing
wind due to orbital motion in a binary system \citep{Solf85}. Another symmetry
axis perpendicular to the jets bisects the outer ``whiskers'' and passes
directly through the CCO. This is highly unlikely to be simply by chance or due
to projection effects, but can be explained if the two objects were in the past
members of a binary system destroyed by a supernova explosion. The observed
25$^{\prime\prime}$ offset between the central star and the CCO is consistent
with this hypothesis. Assuming that the supernova explosion happened
$\sim4.5$\,kyr ago implies that the two stars move away from each other with
a projected velocity of $\sim100$\,km\,s$^{-1}$, which is not unreasonable.

Moreover, the ages of the infrared structure and the SNR shell must be comparable.
Indeed, we already discussed that if the central source is a post-AGB
star it must have entered the mass loss phase quite recently. Furthermore, terminal
wind velocities of massive stars are usually below 1000\,km\,s$^{-1}$,
therefore the 5$^{\prime}$ extension measured for the infrared shell implies a
lower age limit of $\sim4500$\,yr consistent with the existing estimates for
the SNR's age \citep{Yang14}. On the other hand, both the post-AGB phase of the
central star and the X-ray bright phase of the SNR are expected to have
relatively short lifetimes ($\le10000$\,yr), limiting the maximum ages of both
objects. Given that the total lifetime of both stars is much longer than 10\,kyr,
such coincidence is unlikely unless their evolutionary paths are
interconnected, i.e. they were members of the same binary system disrupted by a
supernova explosion $\sim$4-10 kyr ago. In this case, either the supernova
explosion could have been triggered by the enhanced mass transfer from the
companion which entered the AGB phase or vice-versa.

\subsection{Accretion onto the neutron star}
The ``buried field'' scenario proposed by \cite{Ho11} to explain apparently low
magnetic fields of the CCOs invokes a powerful accretion episode soon after the
supernova explosion. Mass loss rates of AGB stars can reach
$\sim10^{-4}M_\odot$ yr$^{-1}$ \citep{Meijerink03_agb_mass_loss}, and although
the current distance between the central star and the CCO implies that the plasma
density in the vicinity of the neutron star is too low to explain the observed X-ray
luminosity of 10$^{34}$\,erg\,s$^{-1}$ by accretion of this material onto the
neutron star, the situation could have been quite different in the past.

Indeed, accretion at much higher rates should have been possible
when the neutron star was closer to its former companion and was thus submerged
in the much denser and slower wind, or even in a common envelope surrounding
the former binary components. In the latter case, accretion rates of up to
$10^{-3}M_\odot$\,yr$^{-1}$ \citep{Chevalier93} can be expected. Indirect
evidence that powerful accretion did indeed take place in the past comes from
the fact that XMMUJ173203.3$-$344518 is the most luminous CCO known to date,
despite not being the youngest. This suggests that it must have also been
heated after the supernova explosion, likely due to accretion. Moreover, as
discussed above, the gap between the outer ``whiskers'' visible in the
24-70\,$\mu$m images is perfectly aligned with the line connecting the central
source and the CCO. This can naturally be explained if the dust was destroyed
by strong X-ray emission from the neutron star during its early high-accretion
rate state. This can also be the reason for the overall asymmetric shape of the
extended emission. The proposed model to explain the observational evidence is
thus that the neutron star was indeed accreting at high rates in the past,
shortly after the supernova explosion. This would imply that the system
exhibits the first observational evidence for the ``buried magnetic field''
scenario proposed by \cite{Ho11} to explain the slow spin evolution of CCOs.

\section{Conclusions} We report on the discovery of the infrared shell
surrounding a post-AGB star projected at the geometrical centre of the
supernova remnant G353.6$-$0.7. Based on the analysis of infrared data, we
conclude that the observed emission comes from a dust shell heated by the
central star likely located within the supernova remnant. Furthermore,
additional dust heating by the CCO suggests that the shell encloses both the
central star and the CCO and, therefore, resides within the supernova remnant.
Based on the morphology of the infrared shell and comparison of its
evolutionary timescale with that of the SNR, we conclude that the post-AGB star
and the progenitor of the remnant's CCO were likely members of the same binary
system disrupted by the supernova explosion. This could be the first evidence
for a binary origin of a CCO. We note that, in this case, accretion of
metal-rich material onto the neutron star early on after the supernova
explosion could explain the slow spin evolution of CCOs via the ``buried
field'' scenario and high carbon content in the atmosphere of the neutron star.

In addition, we estimated the mass of the dust in the shell and concluded that
it significantly exceeds the expected dust yield from the central post-AGB star and
is closer to that from the supernova. This suggests that the interaction of the
central stars' wind with the ejecta enhances the dust formation in central
parts of the remnant. We note that a significant fraction of all core-collapse
supernovae are expected to occur in binary systems, often with medium sized
companions, which have similar evolutionary paths as the central star in
SNR~G353.6$-$0.7. The contribution of such systems to the Galactic dust budget
has not been assessed thus far, and, given the large estimated dust mass around
the central star in SNR~G353.6$-$0.7, could be an important dust formation
mechanism.
 
\section*{Acknowledgements}

The authors thank Emma Whelan and Klaus Werner for their very useful comments.
The authors acknowledge support from Deutsches Zentrum f{\"u}r Luft- und
Raumfahrt (DLR) through DLR-PT grants 50\,OR\,1310 and 50\,OR\,0702, from
Deutsche Forschungsgemeinschaft (DFG) through grants DFG\,WE\,1312/48-1 and
Emmy Noether research grant SA2131/1-1, from Bundesministerium f{\"u}r
Bildung und Forschung (BMBF) through DESY-PT grant 05A14VT1, and from the European Space Agency PRODEX Programme - Contract Number 420090172.
This work is based
in part on archival data obtained with the Spitzer Space Telescope, which is
operated by the Jet Propulsion Laboratory, California Institute of Technology
under a contract with NASA. This work also made use of Hershel and XMM-newton
data. Herschel is an ESA space observatory with science instruments provided by
European-led Principal Investigator consortia and with important participation
from NASA. XMM-Newton is an ESA science mission with instruments and
contributions directly funded by ESA Member States and NASA. This publication
makes use of data products from the Wide-field Infrared Survey Explorer, which
is a joint project of the University of California, Los Angeles, and the Jet
Propulsion Laboratory/California Institute of Technology, funded by the
National Aeronautics and Space Administration. This research has also made use
of the NASA/IPAC Infrared Science Archive (IRSA), which is operated by the Jet
Propulsion Laboratory, California Institute of Technology, under contract with
the National Aeronautics and Space Administration. All infrared data used in
this work is available through IRSA.





\bibliography{databaseu}
\bibliographystyle{mnras}


\appendix

\section{Appendix}
\subsection{Dust mass and thermal balance.}
\label{dustmass}
The infrared flux from the dust envelope observed at Earth can be approximated as
$$
\lambda F_{\lambda}=\frac{2\pi a^2N}{4\pi D^2}Q_{IR}(a,\lambda)\frac{2hc^2/\lambda^4}{(e^{hc/\lambda kT_{\mathrm{dust}}}-1)}
$$
where $D$ is the distance to the source, $Q_{IR}\simeq(2\pi a/\lambda)^\beta$ is
the dust emissivity of grains with size $a$ at a given wavelength
\citep{Draine84} and $N$ is the total number of dust grains. The dust mass
$M_{\mathrm{dust}}=4/3\pi a^3\rho N$ can, therefore, be estimated from the observed flux
if the grain size and density $\rho$ are known or assumed. The problem is, however,
that the emissivity index $\beta$ and grain parameters are generally not known,
which strongly affects the estimated dust temperature and mass. On the other hand,
dust only emits because it is heated by an external source. To within an order
of magnitude, the thermal balance of dust grains in an external radiational
field at given distance $d_*$ from the heating source can be described as
$$
L_*(T_*)/4\pi d_*^2<Q_{UV}(T_*)> = 4<Q_{IR}(T_{\mathrm{dust}})>\sigma_B T_{\mathrm{dust}}^4
$$
where $L_*, T_*$ are its luminosity and effective temperature, $T_{\mathrm{dust}}$ is the
dust temperature, and $<Q_{IR}>$ and $<Q_{UV}>$ are the Planck-averaged dust emission
and absorption cross-sections \citep{Draine84}. 
For a given emissivity law (i.e.
emissivity index $\beta$), the observed dust temperature at a given distance
can be estimated using the ratio of 24 and 70$\mu$m fluxes. On the other hand,
$\beta$ can be constrained using the broadband SED. In fact, the situation is
more complicated due to the dependence of $<Q_{IR}>, <Q_{UV}>$, and $\beta$ on
temperature, so the observed flux ratio must be modelled simultaneously with the
broadband SED to obtain a self-consistent solution.

\begin{figure}
	\centering
		\includegraphics[width=0.49\textwidth]{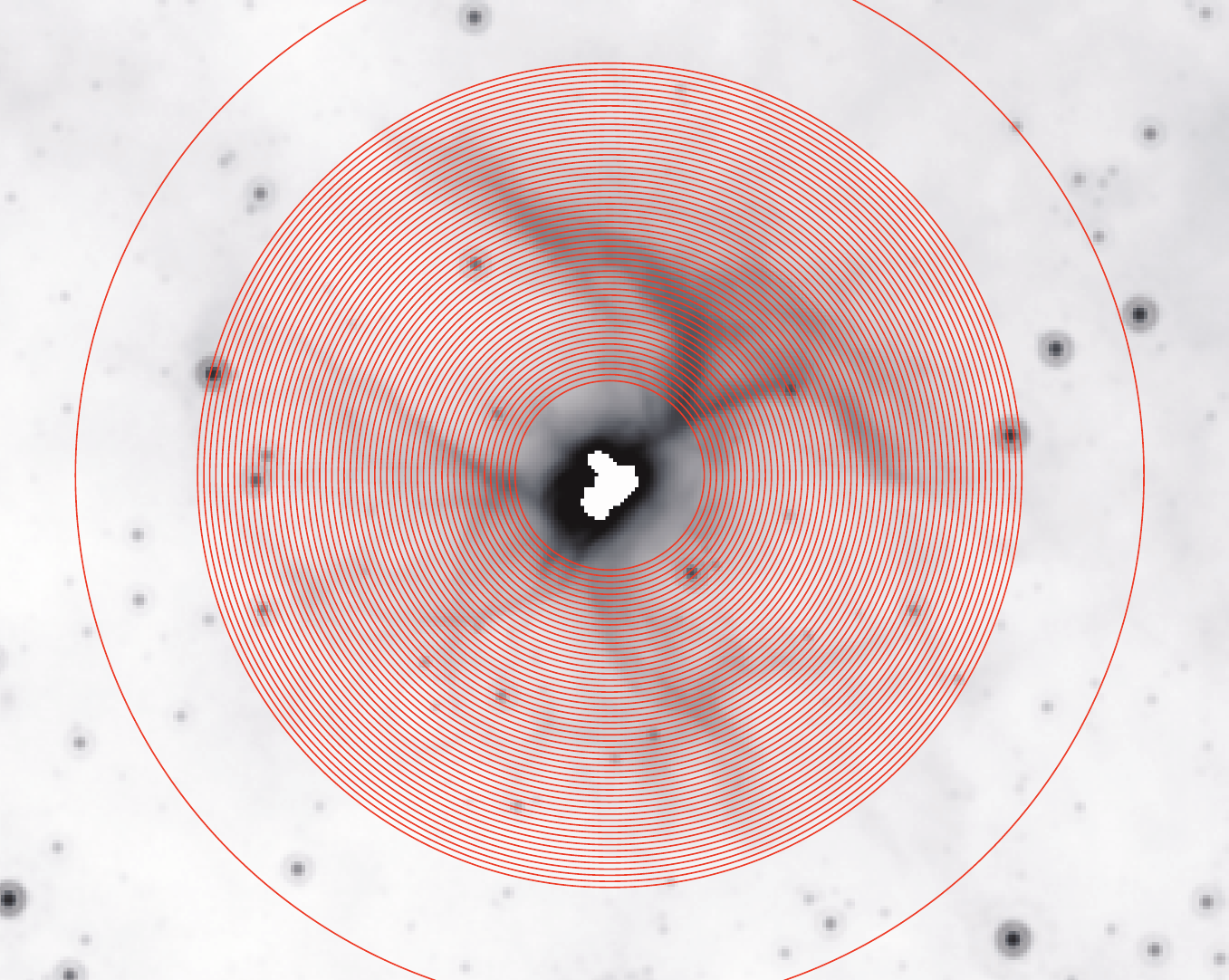}
	\caption{Regions used to measure the radial flux profiles in the 24\,$\mu$m and 70\,$\mu$m bands (the MIPS 24\,$\mu$m image is shown for reference).
	The contribution of point sources was excluded from each image. The outer region was used to estimate the background.}
	\label{fig:s1_ir_reg}
\end{figure}

We measured the radial flux profiles in the 24 and 70$\mu$m bands using a set of
circular annuli with radii from 60$^{\prime\prime}$ to 270$^{\prime\prime}$ and
a width of 4$^{\prime\prime}$, centred on the central source as shown in
Fig.~\ref{fig:s1_ir_reg}. We excluded the inner $1^\prime$ core from the fit
due to saturation of MIPS data and a potentially complex structure due to
presence of the ``jets'' and additional heating of the dust by X-rays from the
neutron star. From the observed fluxes we subtracted the background measured in
the outer annulus with a radius of 270-350$^{\prime\prime}$, where no source
emission is apparent. We also included a systematic error of 5\% to account for
absolute flux calibration uncertainties of both MIPS and PACS and included a
temperature-dependent colour correction as described in the respective
instruments' documentation, to calculate the expected fluxes for a given
temperature. We also assumed a distance to the object from Earth of 3.2\,kpc,
and an astronomical silicate dust composition with cross-sections from
\citep{Draine84,Laor93}. We would like to emphasise here that thermal balance
equation itself does not depend on the assumed distance to the source $D$.
Indeed, the observed luminosity of the central source $L_*\propto F_{\mathrm{obs,bol}}D^2$
has the same dependence on distance as the physical distance from grains to the central
source $d_*^2\propto (\alpha D)^2$. However, the observed bolometric flux
$F_{\mathrm{obs,bol}}$ is uncertain due to strong reddening of the optical star.
Therefore, either extinction coefficient or distance to the source have to be
assumed to constrain the luminosity of the source. We verified also that the main
results are not changed if a carbon composition as provided by the same authors
is assumed. Calculated flux ratios were then compared to the observed values by
simultaneously fitting the flux ratio profile and the SED. Once the parameters
of the fit are fixed, the dust mass in each radial region can be calculated
from the respective normalisation in the same way as for the SED.

Note that we ignored the projection effects as the true geometry of the shell
is not known. Dust can be considered optically thin in the infrared, so fluxes
measured within annular regions represent, in effect, a combination of emission
from individual spherical shells defined by corresponding annular extraction
regions. The contribution of each shell to the flux measured in a given region
is proportional to the volume occupied by dust within the intersection between the
respective spherical shell and annular cylinder. In principle, knowledge of
the three-dimensional structure of the shell is, therefore, required to model the
observed emission. However, a strong dependence of the total flux on
temperature and fast drop of temperature with distance to the heating source
imply that the contribution of outer annuli to the flux measured from a given
region is negligible. Indeed, even without accounting for projection effects
the observed flux ratio in the 24 and $70\,\mu$m bands is consistent with the
predicted radial temperature profile.

However, projection directly affects the dust mass estimate. It is trivial to
show that the volume of intersection between the co-centric spherical shell and
annular cylinder relative to the total volume of the respective spherical shell
is $V_{\mathrm x}/V_{\mathrm shell}=(r_o^2-r_i^2)^{1.5}/(r_o^3-r_i^3)$, where $r_o,r_i$ are the
outer and inner radii of the shell. For regions we used to estimate the
temperature profile, these are $0.3-0.1$, for the inner and outer parts of the
shell, respectively. Together with the dust mass estimated using the flux in
individual regions, this implies that by ignoring projection effects we
underestimate the total dust mass by a factor of $\sim7$. Note, however, that
the observed morphology of the infrared emission implies that the dust shell is
clearly not spherically symmetric, so in reality this factor shall probably
be a factor of two lower, which implies the dust mass of $1.5M_\odot$ for the
preferred solution with grain size of $a = 0.1\,\mu$m.



\subsection{X-ray absorption column and extinction.} 
\label{xraynh}
The expected optical extinction coefficient can be calculated from the X-ray
absorption column as $A_V\simeq N_H/1.6\times10^{21}$ \citep{Vuong03}. Based on
the X-ray spectrum of the CCO, an absorption column of $(1.4-1.9)\times10^{22}$
has been reported by \citep{Klochkov15}, which translates to $A_V\sim8.7-12$ and
agrees well with the value obtained from the SED fit ($A_V=8.67$).

\subsection{X-ray emission from the central star.} The central source is
detected in X-rays as well (see Fig.~\ref{fig:closeup}). Based on the
\emph{Chandra} X-ray spectrum, its flux is estimated to
$\sim10^{-13}$\,erg\,cm$^{-2}$\,s$^{-1}$, corresponding to a luminosity of
$L_x\sim10^{32}$\,erg\,s$^{-1}$ (adopting the absorption column of
$N_H\sim1.4\times10^{22}$ \citep{Klochkov15} and a blackbody spectrum with a
best-fit temperature of $\sim3\times10^6$\,K). We omitted this measurement in the
SED modelling since the hard X-ray emission of X-ray active stars is believed to
be non-photospheric, but rather due to either enhanced coronal activity or
strong shocks in stellar winds. The ratio of the X-ray to optical luminosity is
consistent with values typically observed from X-ray active AGB stars
\citep{Ramstedt12_xray_to_optical}. Thus, we considered the detection of X-rays
from the central source as an additional argument to support the classification
of IRAS~17287-3443 as a post-AGB star.

\bsp	
\label{lastpage}
\end{document}